\documentclass[twocolumn,amsmath,amssymb,showpacs,prb]{revtex4}

\usepackage{graphicx}
\usepackage{color}
\usepackage{soul}
\usepackage{psfrag}
\usepackage{epsfig}
\usepackage{bbm}
\usepackage{bm}
\usepackage{ifthen}
\usepackage{calc}
\usepackage[latin1]{inputenc}
\usepackage[T1]{fontenc}
\usepackage{amsfonts}
\usepackage{amstext}
\usepackage{pst-all}
\usepackage[english]{babel}
\usepackage{subfigure}
\usepackage{rotating}
\usepackage{textcomp}
\usepackage{color,hyperref}

\newcommand{\bld}[1]{\boldsymbol #1}
\newcommand{\ket}[1]{| #1 \rangle}
\newcommand{\bra}[1]{\langle #1 |}
\newcommand{\rb}[1]{\left( #1 \right)}

\newcommand{\ew}[1]{\langle #1 \rangle}
\newcommand{\beq}{\begin{eqnarray}}
\newcommand{\eeq}{\end{eqnarray}}

\newcommand{\op}[2]{| #1 \rangle \langle #2 |}

\newcommand{\eq}[1]{Eq.~(\ref{#1})}

\begin{document}

\title{Two-Particle Dark State in the Transport through a Triple Quantum Dot}
\author{Christina Pöltl, Clive Emary and Tobias Brandes}
\affiliation{
  Institut für Theoretische Physik,
  TU Berlin,
  Hardenbergstr. 36,
  D-10623 Berlin,
  Germany
}

\date{\today}

\begin{abstract}
We study transport through a triple quantum dot in a triangular geometry
with applied bias such that both singly- and doubly- charged states
participate.  We describe the formation of electronic dark states ---
coherent superpositions that block current flow --- in the system, and
focus on the formation of a two-electron dark state.  We discuss the
conditions under which such a state forms and describe the signatures that
it leaves in transport properties such as the differential conductance and
shotnoise.
\end{abstract}
\maketitle

\section{Introduction}
Dark states (DSs) are a quantum-mechanical phenomenon originally discovered as a dark line in the fluorescence of sodium atoms \cite{{dark1}}.
For a particular configuration of atomic transitions and driving fields  \cite{dark1,dark2, dark3}, relaxation can drive the atomic electron into a superposition state that is completely decoupled from the light field --- this is the dark-state.

The concept of the DS has been generalized to mesoscopic transport \cite{010,011,sie,Fao,003,004,005}. In contrast to quantum optics, such systems are connected to electron reservoirs, and enable one to study the influence of DSs on non-equilibrium transport properties such as current, shotnoise \cite{001}, and full counting statistics \cite{002}.
The first transport DS was proposed in a system similar to those of quantum optics, in which microwave fields were used to create a DS in a double quantum dot \cite{{010},{011}}. Subsequently, an all-electronic mechanism for the creation of DS has been described \cite{{003},{004},{005}}, in which coherent tunneling of electrons plays the role of the (classical) driving fields.
The analysed system was a triple quantum dot (TQD) in a triangular geometry in the strong Coulomb blockade regime.

In the afore-mentioned systems, the DS is a single-electron state formed by interaction with the environment.  In the current work, we investigate the effects of a second electron on DS formation.
This we do in a mesoscopic transport context, where a change of the chemical potential of the reservoirs can simply lead to the inclusion of two-electron states in the transport window.  In particular we study the influence of two-electron states on the transport through a TQD.
The inclusion of doubly-charged states is particularly important because the formation of the original single-electron DS depends not just on destructive interference, but also on the strong Coulomb Blockade.
The inclusion of two-electron states might therefore be expected to inhibit the appearance of DSs in such systems.
However, as we will show, dark-state formation is in fact possible with two electrons, but only under certain circumstances.  Even if the formation of the two-electron DS is incomplete, partial dark states still leave obvious signatures in the transport properties such as negative differential conductance and superPoissonian shotnoise.

This work comes against a backdrop of growing theoretical interest in the transport properties of triple quantum dots \cite{TQD1,TQD2,TQD3,TQD4}, and in the suppression of current due to interference phenomena \cite{Inf1,Inf2,Inf3} as distinct from other current blockade mechanisms in quantum dots such as Coulomb \cite{Blo3}, spin \cite{Blo2}, isospin \cite{Blo1} and Franck-Condon \cite{Blo4} blockades.
This work also has experimental relevance as a number of groups have published results of transport measurements on TQDs \mbox{\cite{{E01},{E02},{E03},{E04},{E05}}}.  The finite-bias calculations that we present here should facilitate the experimental investigation of dark-state effects such as the break up of Coulomb blockade diamonds in the stability diagram of the TQD due to one- and two-electron DSs.

\section{Model}

\begin{figure}[top]
 \centering
\subfigure[Two electrons in the same dot]{
\includegraphics[width=0.4\textwidth]{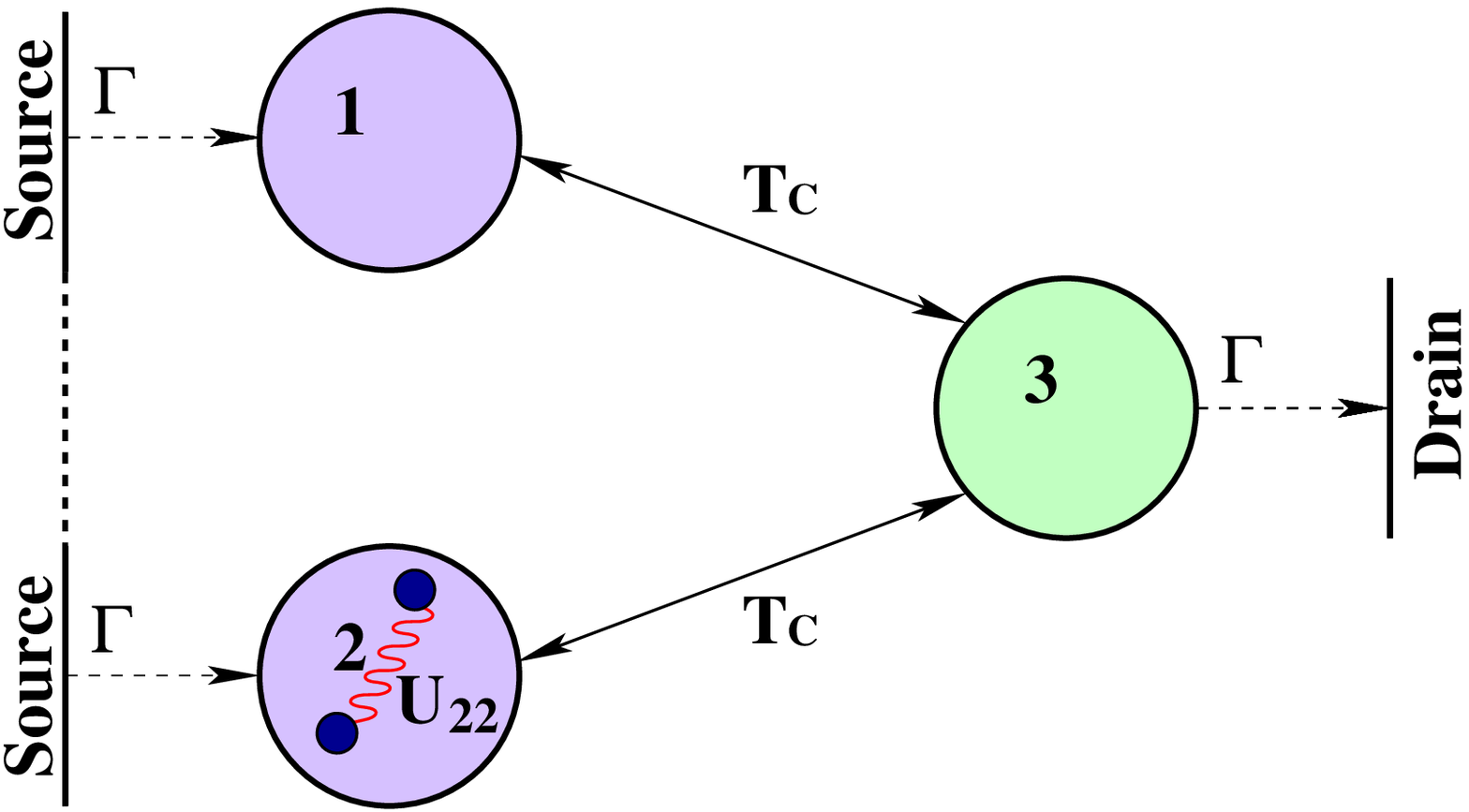}}
\subfigure[Two electrons in different QDs]{
\includegraphics[width=0.4\textwidth]{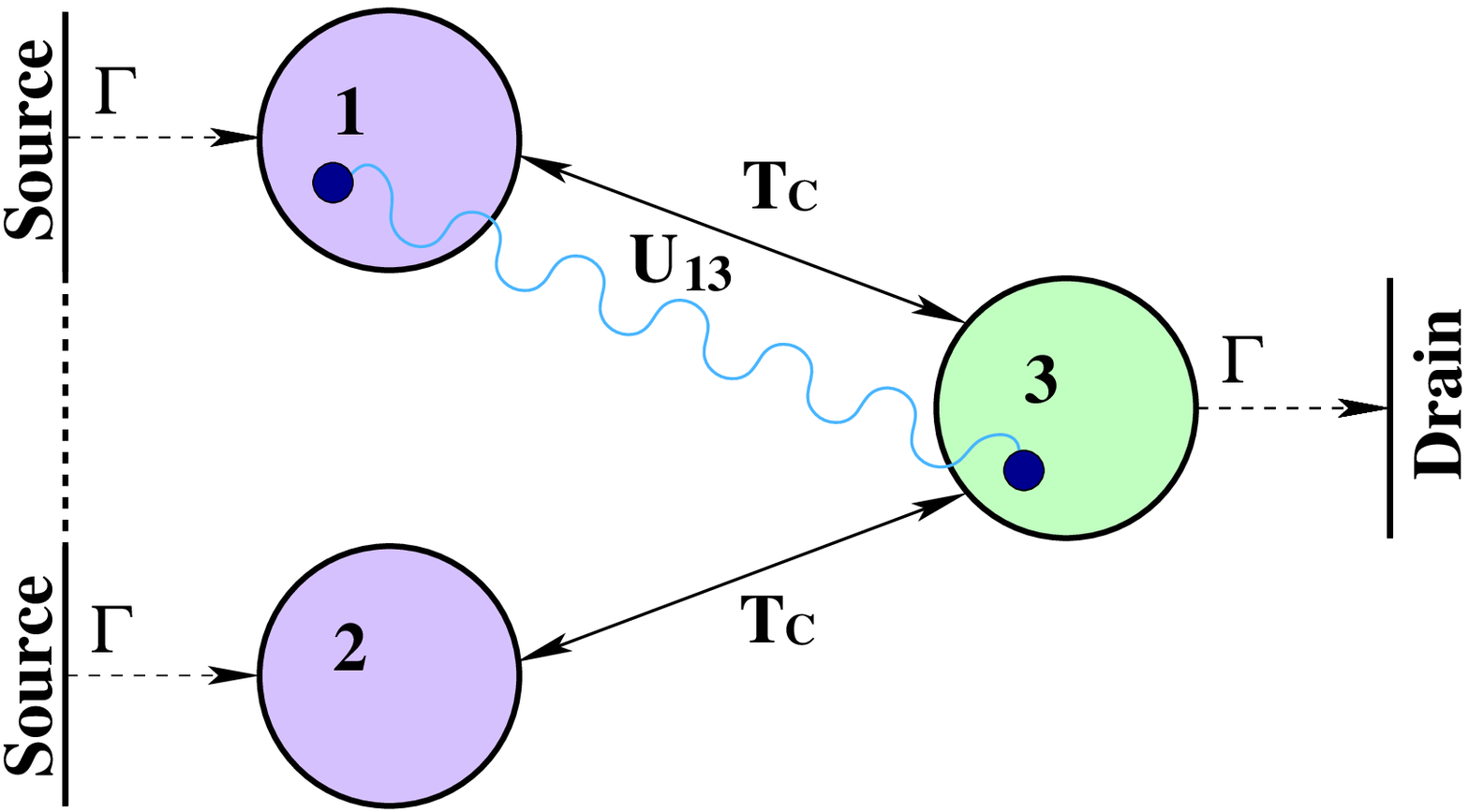}}
\caption{
The triple quantum dot (TQD) in triangular geometry is connected to two sources and one drain. In the infinite bias limit, the rate at which electrons enter and leave the TQD is $\Gamma$, the same for all three leads.
The QDs are coupled to each other by the coherent tunnel amplitudes $T_C$ as shown. Zero, one or
two electrons are allowed to be in the TQD at a time. The charging energy between two electrons in the same QD e.g. (a)  $U_{22}$ is, in general, larger than for two electrons in different QDs, e.g. (b) $U_{13}$.  }
\label{fig:System2}
\end{figure}

Our system consists of three quantum dots (QDs) arranged in a triangular geometry with a single relevant orbital level in each dot (see Fig.~\ref{fig:System2}). 
QD1 and QD2 are connected to electron source reservoirs and QD3 is connected to a drain reservoir.
In the infinite bias limit, the rate at which electrons enter and leave the TQD is $\Gamma$, which we assume the same for all three leads.  The levels in QD1 and QD2 are coupled coherently to QD3 with a tunnel amplitude $T_C$.

We assume that the system is in the Coulomb blockade regime, and we adjust the lead chemical potentials such that the only relevant charge states have zero, single and double excess electrons. We incorporate the single-electron charging energies into the energies of single electron orbital levels. With the addition of a second electron we associate an additional charging energy $U_{ij}$, with  $i \leq j=1,2,3$, describing the locations of the two electrons.

\subsection{Hamiltonian}
The system Hamiltonian $\hat{H}_D$ of the closed TQD is a Hubbard-type model with three sites 
\begin{align}
\hat{H}_{\text{D}} \!=&\! \sum_{i \sigma}\!E_{i}\hat{n}_{i \sigma}\!+\!T_C\sum_{\sigma}(d^{\dagger}_{1 \sigma}d_{3 \sigma}\!+\!d^{\dagger}_{2 \sigma}d_{3 \sigma}\!+\!H.c.) + \nonumber \\
&\!   \sum_{i}^3 U_{ii} \hat{n}_{i \uparrow}\hat{n} _{i\downarrow}+ \sum_{i,j,i<j}^3 \sum_{\sigma \sigma'} U_{ij} \hat{n}_{i \sigma} \hat{n}_{j \sigma'}, \label{QD_m1} 
\end{align}
where $d_{i\sigma}$ is the annihilation operator, and $\hat{n}_{i\sigma}$ is a corresponding number operator of an electron in dot $i$ with spin $\sigma$. We assume spin-independent energy levels and denote the energy of the single-electron level of the QD $i$ as $E_{i}$.    
Separating the Hamiltonian $\hat{H}_D$ into one-and two-electron parts, 
\begin{eqnarray}
\hat{H}_{D} &=& H_\uparrow + H_\downarrow + H_{\uparrow \uparrow} + H_{\downarrow \downarrow} + H_{\uparrow\downarrow}, 
\end{eqnarray}
we explicitly construct the corresponding matrices in a basis of relevant many-body states.  
With the single-electron basis $\left\{\ket{1\sigma},\ket{2\sigma},\ket{3\sigma}\right\} $, the Hamiltonians $H_\uparrow$ for a single spin-up electron in the TQD and $H_\downarrow$ for a spin-down electron in the TQD are 
\begin{eqnarray}
H_\uparrow=H_\downarrow=\begin{pmatrix}
E_1                &0 &T_C\\
0	&E_2                &T_C \\
T_C              &T_C             &E_3   \end{pmatrix}
.
\end{eqnarray}
The parallel-spin  Hamiltonians $H_{\uparrow \uparrow}$ and $H_{\downarrow \downarrow}$ in the two-electron basis $\left\{\ket{1\sigma2\sigma}, \ket{1\sigma3\sigma}, \ket{2\sigma3\sigma}\right\}$ read
\begin{eqnarray}
H_{\uparrow\uparrow} \!= \! H_{\downarrow\downarrow} \! = \! \begin{pmatrix}
E_1\!+\!E_2\!+\!U_{12} &T_C                &-T_C              \\
T_C           &E_1\!+\!E_3\!+\!U_{13}      & 0 \\
-T_C           &0 &E_2\!+\!E_3\!+\!U_{23}     \end{pmatrix}.
\end{eqnarray}
The opposite-spin two-electron Hamiltonian $H_{\uparrow\downarrow}$ is a $9\times9$ matrix, which is given in Appendix~\ref{appen_1}. We denote the basis of the above matrices the  \textquotedblleft localized basis\textquotedblright.  In subsequent calculations we set all 
$E_{i }=E_0+E_\text{Gate}=-e V_\text{Gate}$ (in the infinite bias calculations we set $E_\text{Gate}=0$).

A key parameter in our discussion of the two-particle dark state will be the
difference between the charging energy $U_{11}$ and the charging energy $U_{12}$:  
\beq
  \delta_U \equiv U_{11}-U_{12},
\eeq
which we will call the ``charging-energy difference''. 
Since two electrons in a single QD are closer than two electrons in different QDs, one expects that the charging energies between electrons in different QDs ($U_{ij}$; $i\neq j$) will be smaller than those for electrons in the same QD  ($U_{ii}$) and thus we focus here on $ \delta_U \ge 0$, (although note, for example, Ref.~\cite{nU1}). 
We will discuss two sets of charging energies: Firstly a highly symmetric situation where all $U_{ii}$ are equal and all $U_{ij}$ for $i\neq j$ are also equal. Although this is the simplest situation, its high symmetry leads to non-generic features as we will see. We therefore consider a second set of charging energies with $U_{11}=U_{22}\neq U_{33}$ and $U_{12} \neq U_{13} = U_{23}$, which breaks the symmetry and leads to more typical results.

The TQD is connected to three electron reservoirs that are described with the Hamiltonian
\begin{eqnarray}
\hat{H}_{\text{res}} &=& \sum_{i=1,2,3} \sum_{k,\sigma} \epsilon_{ik} c^\dagger_{i k\sigma} c_{ik \sigma}, \label{QD_m3}
\end{eqnarray}
where  $i$ labels the reservoirs ($1,2 =$ source, $3=$drain).
The TQD and the reservoirs are connected by the tunnel Hamiltonian 
\begin{align}
\hat{H}_{\text{T}} =& \sum_{i=1,2,3} \sum_{k,\sigma}  V_{ik}c^{\dagger}_{ik\sigma}d_{i \sigma} +\mathrm{H.c.}
\label{QD_m4}
.
\end{align}
We assume spin-independent reservoir energies $\epsilon_{ik}$ and tunneling amplitudes $V_{ik}$.

\subsection{Method}
We use two different approaches to calculate the current and the Fano factor of the TQD based on the sequential-tunneling (i.e. second order in tunneling Hamiltonian, $H_T$) Master equation\cite{{F01},{N01},{N02}}.  The two approaches are:

\subsubsection{The generalized master equation}
In the infinite-bias limit, the second-order Born-Markov master equation 
is believed to be exact for flat bands provided that coherences between system states are explicitly included   \cite{{Gur},{Gur2}}. 
In the localised basis, this approach leads to the generalized master equation (GME). We assume that the chemical potentials are positioned far from other relevant energies, $|\mu_i|> U_{ij} , T_C, V_{ik}, E_i$, but such as to exclude three-electron states and states in which an electron occupies an excited orbital state of the QDs. 
We use the TQD Hamiltonian $H_D$ and approximate
the chemical potential of the sources to be $\mu_{1,2}=\infty$ and the chemical potential of the drain to $\mu_3=-\infty$.
In the GME approach, the density matrix $\rho(t)$ contains entries for the populations of the empty state, and one- and two-electron states in the localized basis.  Furthermore, $\rho(t)$ also contains all coherences within each charge sector.
Within the GME, the time-evolution of the density matrix is then given by the Lindblad form \cite{lin}
\begin{align}
\frac{d\rho}{dt} \!  = & -i[\hat{H}_{\text{D}},\rho]\! 
+\!\!\! \!\!\! \!\sum_{\sigma; j=1,2} \! \!\!\! \Gamma_{j\sigma} \!\! \left(d^\dagger_{j \sigma} \rho d_{j \sigma} \!-\!\frac{1}{2}d_{j \sigma} d^\dagger_{j \sigma}\rho\!-\!\frac{1}{2}\rho d_{j \sigma} d^\dagger_{j \sigma}\right) \nonumber \\ 
&+ \!\sum_{\sigma}\Gamma_{3\sigma} \left(d_{3\sigma}\rho d^\dagger_{3\sigma}\!-\!\frac{1}{2}d^\dagger_{3\sigma} d_{3\sigma}\rho\!-\!\frac{1}{2}\rho d^\dagger_{3\sigma} d_{3\sigma}\right)
\label{auh}.
\end{align}
The second quantized operators appearing in this equation are given in the relevant many-body basis in Appendix \ref{appen_2}.
We assume energy- and spin-independent tunnel rates
\begin{eqnarray}
 \Gamma_{i \sigma} =2\pi  \sum_{k}  |V_{ik}|^2 \delta(\epsilon-\epsilon_{ik}),
\end{eqnarray}
with $i=1,2,3$. In the following we set all $\Gamma_{i\sigma} =\Gamma$.
The Lindblad master equation can then be written in super-operator formalism where the density matrix $\bld{\rho}$ is written as a vector. 
The equation of motion is then
\begin{eqnarray}
\dot{\bld{\rho}}(t)=M\bld{\rho}(t), \label{master_b} 
\end{eqnarray} 
where $M$ is the Lindblad super-operator.

The stationary density matrix of the system is $\bld{\rho}_0=\lim_{t\rightarrow\infty}\bld{\rho}(t)$ and in practice this is determined as the null\-vector of $M$, which is unique here.   For our TQD, the steady-state current is then given by 
\begin{align} 
\ew{I}=&e \sum_{\sigma}  \Gamma_{3 \sigma} \big(
\bra{3\sigma} \bld{\rho_0} \ket{3\sigma}+ 
2 \bra{3\!\!\uparrow\!3\!\!\downarrow\!\!} \bld{\rho_0} \ket{3\!\!\uparrow\!3\!\!\downarrow} \nonumber \\
& +\bra{1\sigma3\sigma} \bld{\rho_0} \ket{1\sigma3\sigma}+
\bra{2\sigma3\sigma} \bld{\rho_0} \ket{2\sigma3\sigma}\nonumber \\
& + \bra{1\sigma3\bar{\sigma}} \bld{\rho_0} \ket{1\sigma3\bar{\sigma}}+
\bra{2\sigma3\bar{\sigma}} \bld{\rho_0} \ket{2\sigma3\bar{\sigma}} \big).\label{eq_cur}
\end{align}

\subsubsection{Rate equation in the diagonalised basis } 

An alternative approach is to first diagonalise the system Hamiltonian and then write down a rate equation for the populations of the system eigenstates \cite{Breuer}.  We describe this approach as the diagonalised master equation (DME). Unlike the GME, this approach is not restricted to the infinite bias limit, but the disadvantage is that some of the effects of coherence on the transport dynamics are lost.
In the DME approach the TQD Hamiltonian $H_D$ is first written in diagonal form 
\begin{eqnarray}
\hat{H}_{D} &=& \sum_{k=0}^N \epsilon_{k}\op{\beta_k }{\beta_k }, \label{Ham_dia}
\end{eqnarray}
where $\epsilon_k$ is the energy of eigenstate $\ket{\beta_k}$, and $N=21$ is the dimension of $\hat{H}_D$ without the empty state.
The density matrix $\rho_D(t)$ in the DME approach contains only the populations of states $\ket{\beta_k}$ and its time-evolution is given by the rate equation
\begin{eqnarray}
  \dot{\bld{\rho}}_D(t)=W \bld{\rho}_D(t)
  ,
  \label{Dia_master} 
\end{eqnarray}
with elements of the rate matrix given by 
\begin{eqnarray}
	W_{k' k }&=&\sum_{\sigma,i}\Gamma_{i\sigma} \Big\{
	f(\Delta_{k'k}+\mu_i)|\bra{\beta_k }d_{i \sigma}^{\dagger} \ket{\beta_{k'} }|^2  \nonumber \\
	& & 
	+f(\Delta_{k'k}-\mu_i)|\bra{\beta_k } d_{i \sigma} \ket{\beta_{k'} }|^2  
	\nonumber \\
	& & 
	-\sum_{l}^N
	\big(f(\Delta_{lk}-\mu_i)|\bra{\beta_l } d_{i \sigma}^{\dagger} \ket{\beta_k }|^2
	\nonumber \\
	& & 
	+
	f(\Delta_{lk}+\mu_i)|\bra{\beta_l } d_{i \sigma} \ket{\beta_k }|^2 \big) \delta_{\beta_k , \beta_{k'} } \Big\}
	,
	\nonumber
\end{eqnarray}
where $d_{i \sigma}$ is the tunnel operator of the previous section, $f(x)=(1+e^{x/k_B T})^{-1}$ is the Fermi function (with the temperature $T$ and the Boltzmann constant $k_B$), and $\Delta_{k'k}=\epsilon_{k'}-\epsilon_{k}$ is the Bohr frequency of the transition from $\ket{\beta_k }$ to $\ket{\beta_k' }$.

In this case, the steady-state current is given by
\begin{align} 
\ew{I} =& e \Gamma_{3\sigma} \sum_{k',k=0}^N \Big[
 f(\Delta_{k'k}+\mu_3)|\bra{\beta_k }d_{3 \sigma}^{\dagger} \ket{\beta_{k'} }|^2 (\bld{\rho}_{D0})_k  \nonumber \\
&
- f(\Delta_{k'k}-\mu_3)|\bra{\beta_k } d_{3 \sigma} \ket{\beta_{k'} }|^2  (\bld{\rho}_{D0})_k\Big] , \label{eq_cur_dia}
\end{align} 
with $(\bld{\rho}_{D0})_k$ as the $k$th element of the diagonalized steady-state density matrix.

In the infinite bias limit, the Fermi functions of the sources leads are
${
\lim_{\mu_j\rightarrow\infty} f(\Delta_{k'k}-\mu_j)\!=\!1}$ and
${\lim_{\mu_j\rightarrow\infty} f(\Delta_{k'k}+\mu_j)\!=\!0}$, ${j=1,2}$; and
${\lim_{\mu_3\rightarrow-\infty} f(\Delta_{k'k}+\mu_3)\!=\!1}$, and
${\lim_{\mu_3\rightarrow-\infty} f(\Delta_{k'k}-\mu_3)\!=\!0}$
for the drain.
In this limit, the current is then 
\begin{align} 
\ew{I} &= e \Gamma_{3\sigma} \sum_{k,k'=0}^N |\bra{\beta_k }d_{3 \sigma}^{\dagger} \ket{\beta_{k'} }|^2 (\bld{\rho}_{D0})_k 
. \label{eq_cur_dia_inf}
\end{align}

\section{Stationary Transport in the infinite bias limit}

We first consider the infinite bias limit.
It is in this limit that previous calculations on the single-electron TQD have been performed \cite{{003},{004},{005}} and where we expect the GME to be exact.  We therefore discuss the GME results first and then return to a comparison of GME and DME calculations in this limit. 
\begin{figure}[tb]
 \centering
\subfigure[$U_{22}=U_{33}=U_{11}$ and $U_{13}=U_{23}=U_{12}$]{
\includegraphics[width=0.45\textwidth]{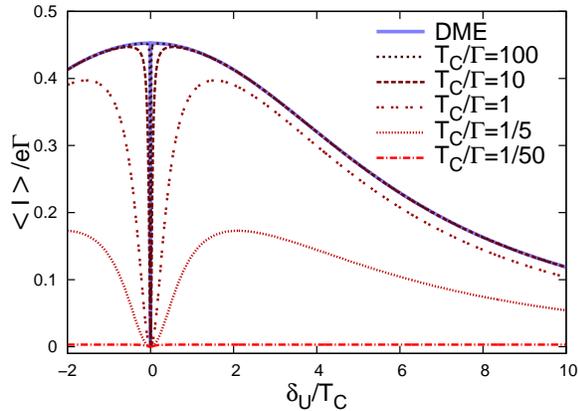}\label{cur_U-Vsym}}
 \subfigure[$U_{11}=U_{22}$, $U_{12}/T_C=10 $, $U_{33}/T_C=15$ and $U_{13}/T_C=U_{23}/T_C=11 $]{
\includegraphics[width=0.45\textwidth]{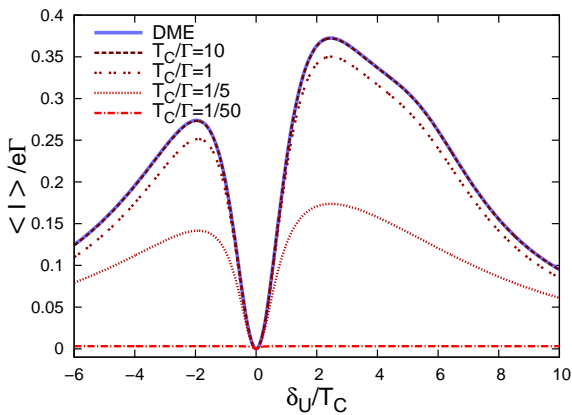}\label{cur_U-Vasym}}
\caption{
	Stationary current $\ew{I}$ through the TQD as a function of the charging-energy difference scaled with the tunnel amplitude $\delta_U/T_C$ for different ratios of $T_C/\Gamma$ and for two choices of charging energies. Results are shown for both GME and DME calculations (the DME calculations are independent of the ratio $T_C/\Gamma$) and both calculations show that at $\delta_U=0$, the current is zero. This is attributed to the formation of the two-electron dark state. In Fig.~\ref{cur_U-Vsym} (symmetric charging energies) the GME calculation shows an antiresonance of finite width about the DS, whereas in the DME calculation the DS appears as a point of discontinuity. For asymmetric charging energies, Fig.~\ref{cur_U-Vasym}, both calculations show an antiresonance of finite width about the DS.
  }
\label{cur_U-V}
\end{figure}

\subsection{Stationary Current}

Figure~\ref{cur_U-V} shows the stationary current $\ew{I}$ as a function of the charging-energy difference $\delta_U$, which we scale with the tunnel amplitude $T_C$.  At the point where the charging-energy difference vanishes ($\delta_U=0$) the current is zero. At this point the system is trapped in the two-electron DS
\begin{eqnarray}
 \rho_{\text{dark}} &=& |\varPsi_{0}\rangle \bra{\varPsi_{0}},
\end{eqnarray}
with
\begin{eqnarray}
 |\varPsi_{0}\rangle=\frac{1}{2}\rb{d^\dag_{1\uparrow}-d^\dag_{2\uparrow}}\rb{d^\dag_{1\downarrow}-d^\dag_{2\downarrow}}\ket{0}. \label{eq_dark_state}
 \label{Psi0}
\end{eqnarray}
For $\delta_U=0$, $|\varPsi_{0}\rangle$ is an exact eigenstate of the two-electron Hamiltonian. As since this state has no occupation on QD3 the electrons in the TQD cannot leave the TQD to the drain and no further electrons can enter the TQD due to the Coulomb blockade, this state is completely dark with zero current.
This situation is analogous to the single-electron DS but here we have two electrons in forming a product state of a spin-up and spin-down single-particle DSs.

\begin{figure}[top]
 \centering
\includegraphics[width=0.45\textwidth]{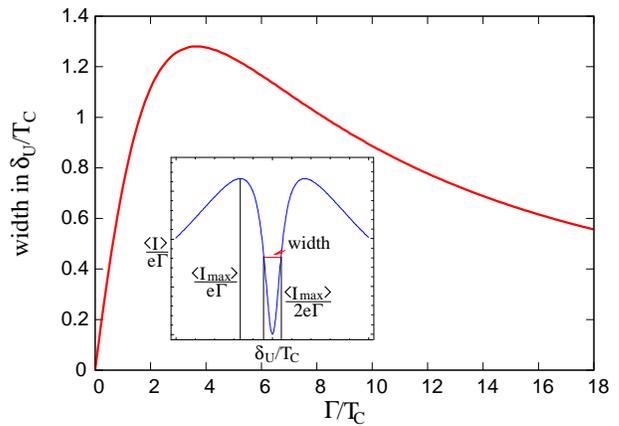}
\caption{Gap width in $\delta_U/T_C$ of the antiresonance at the DS as a function of tunnel rate per tunnel amplitude  $\Gamma/T_C$ for the GME approach with the parameters of Fig.~\ref{cur_U-Vsym}. The size of the gap is given by the width of the antiresonance when the current has half the size of the maximum $\frac{\ew{I_{\text{max}}}}{2(e\Gamma)}$.}
		\label{Gap}
\end{figure}

The current in the GME approach for $\delta_U\neq0$ has a dark resonance profile with a clear antiresonance.  The width of this  antiresonance as a function of the tunnel rate is shown in Fig.~\ref{Gap}.
For small ratios of $\Gamma/T_C<1 $ the gap increases  linearly  from $\Gamma/T_C=0$ (corresponding to the limit of $T_C\rightarrow\infty$ where the DME current and the GME current coincide).  For higher ratios of $\Gamma/T_C$ the size of the gap reaches a maximum at $\Gamma/T_C\approx3.7$, and then decreases.

\begin{figure}[top]
 \centering
\subfigure[$U_{22}=U_{33}=U_{11}$ and $U_{13}=U_{23}=U_{12}$]{
\includegraphics[width=0.45\textwidth]{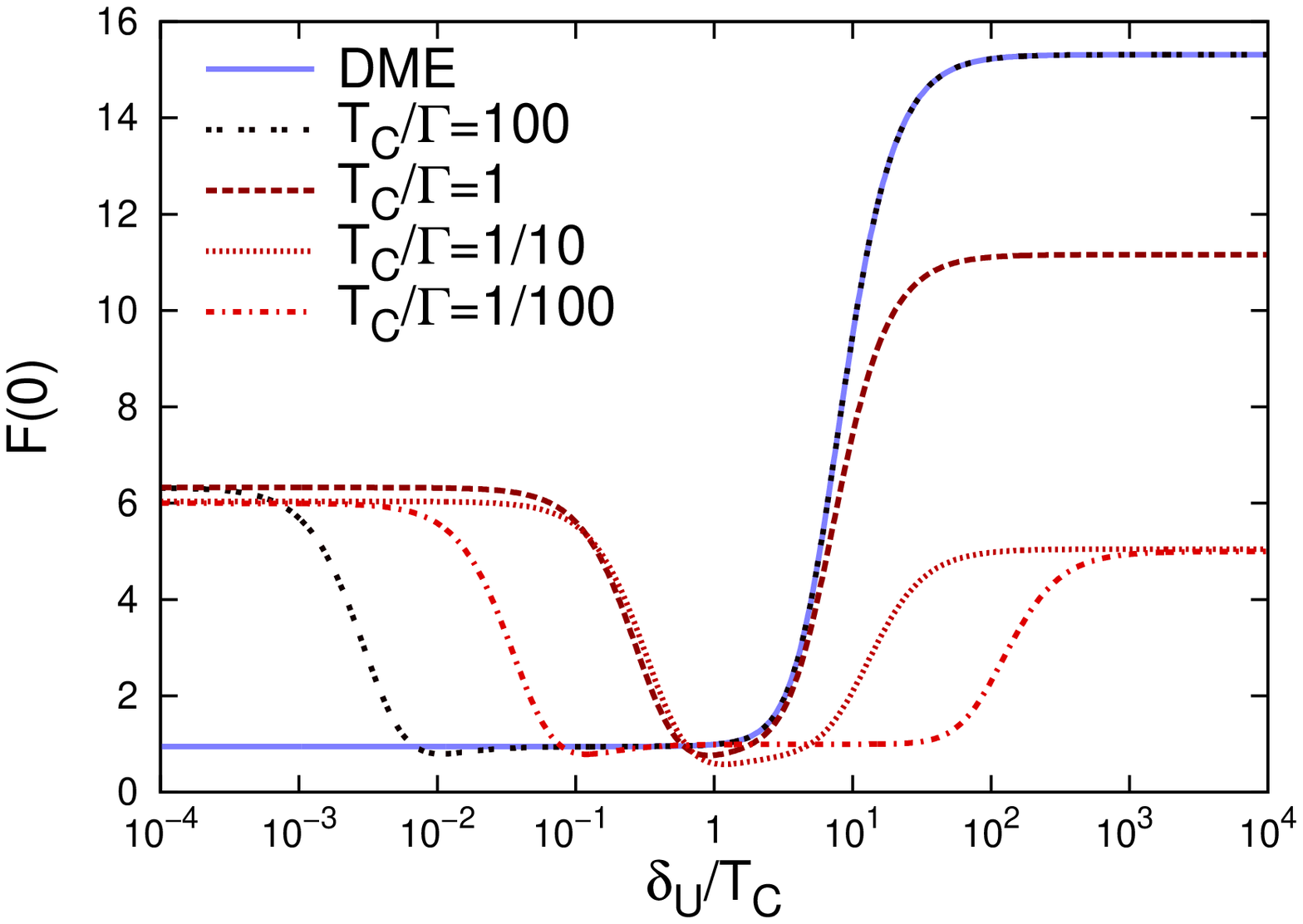}\label{Fano_U-Vsym}}
\subfigure[$U_{11}=U_{22}$, $U_{12}/T_C=10 $, $U_{33}/T_C=15$ and $U_{13}/T_C=U_{23}/T_C=11 $]{
\includegraphics[width=0.45\textwidth]{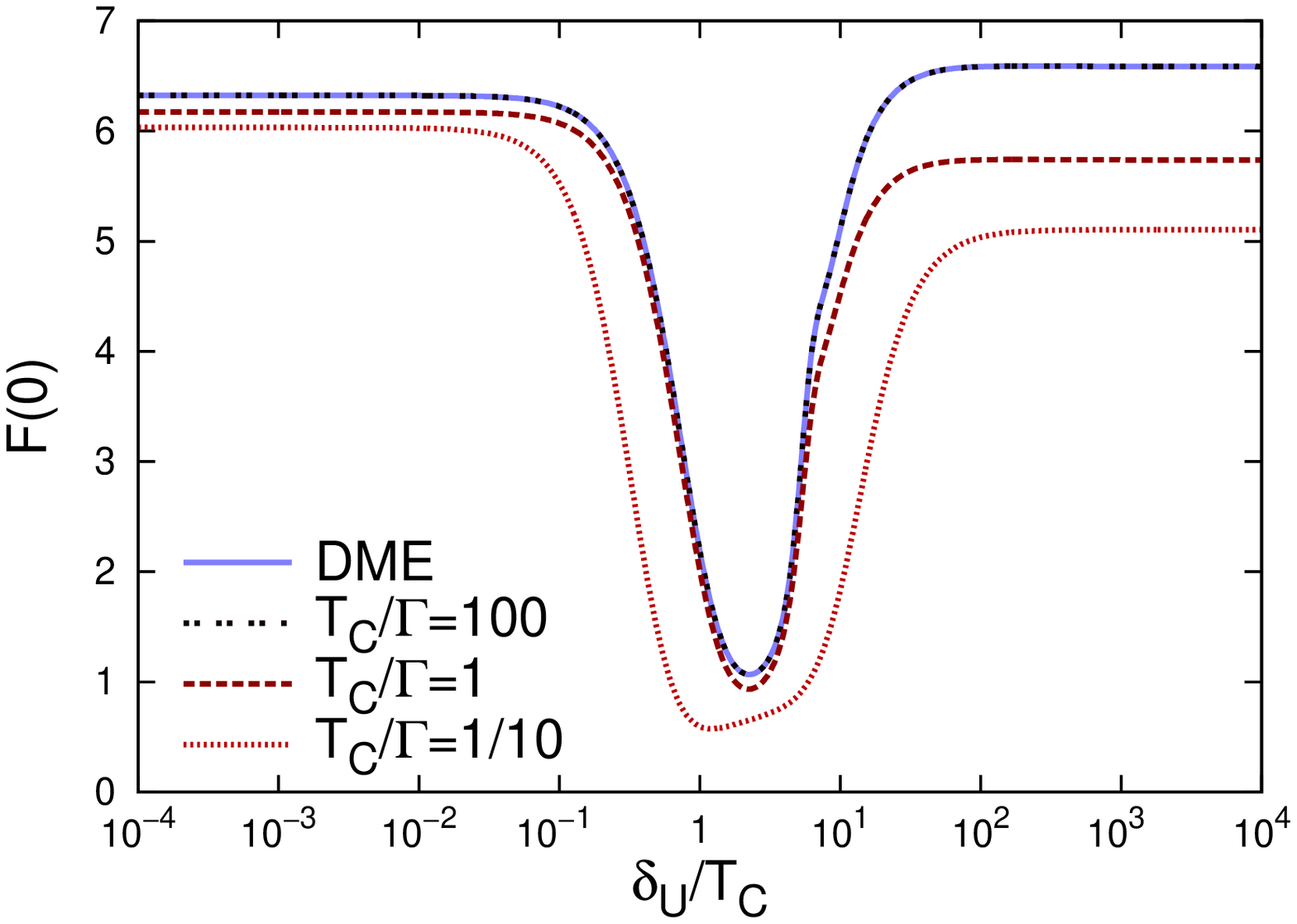}\label{Fano_U-Vasym}}
\caption{
The Fano factor $F(0)$ as a function of the charging-energy difference with results for both GME and DME calculations.  From the GME calculation
for both symmetric (Fig.~\ref{Fano_U-Vsym}) and asymmetric (Fig.~\ref{Fano_U-Vasym}) charging energies,
we observe that the shotnoise is highly super-Poissonian ($F(0)>1$) both near the dark state  $\delta_U/T_C\sim 0$ and in the limit of   $\delta_U/T_C \gg 1$. 
In the DME approach, only the asymmetric case shows super-Poissonian behaviour near the dark state.  The Fano factor in Fig.~\ref{Fano_U-Vsym} is Poissonian for  $\delta_U/T_C < 1$. 
}
\label{Fano_U-V}
\end{figure}

\subsection{Shotnoise}

In the infinite-bias limit, we can write down the $n$-resolved master equation 
\begin{eqnarray}
  \dot{\bld{\rho}}^{(n)} = M_{J}\bld{\rho}^{(n-1)} + M_{0}\bld{\rho}^{(n)} 
\end{eqnarray}
for $\bld{\rho}^{(n)}$, the partial density matrix of the system after $n$ electrons have passed though the TQD \cite{cook81}. Here the total Liouvillian $M$ has been decomposed into two parts: $M_0$, which describes the evolution of system without electron transfer to the drain, and $M_J$, the \textquotedblleft jump operator\textquotedblright, which transfers an electron to the drain.
The matrix elements of the jump operator are obtained from 
\begin{eqnarray}
  M_{J}\bld{\rho} =\sum_{\sigma=\! \uparrow\!, \downarrow\!} \Gamma_{3\sigma} d_{3\sigma}\rho d^\dagger_{3\sigma}
\end{eqnarray}
in the GME approach, and as
\begin{eqnarray}
\rb{M_{J}}_{k'k}= \sum_{\sigma=\! \uparrow\!, \downarrow\!} \Gamma_{3\sigma} |\bra{\beta_k }d_{3 \sigma}^{\dagger} \ket{\beta_{k'} }|^2 
\end{eqnarray}
in the infinite-bias limit of the DME.

The advantage of the $n$-resolved master equation is that it can be used calculate not only the stationary current (given by $ \ew{I}=e \: \text{Tr} \left(  M_J \bld{\rho}_0 \right)$ in this language), but also the full counting statistics of the current \cite{F01}.  Here we concentrate on the shotnoise \cite{001}, and from the method used in Ref.\cite{002}, the zero-frequency Fano factor $F(0)$ can be expressed as 
\begin{eqnarray}
  F(0)=1-\frac{2}{\langle I\rangle}\sum_{k=1}^{N_M} \frac{C_k}{\lambda_k}, 
\end{eqnarray}
with $C_k= \sum_{\alpha=0}^{N}\sum_{\beta,\gamma,\mu=0}^{N_M}(M_J)_{\alpha\beta}v_{\beta k}(v_{k\gamma})^{-1}(M_J)_{\gamma\mu}(\rho_{0})_\mu$ and $\lambda_k$ the eigenenergies ($\lambda_0=0$). 
Here, $(M_J)_{\alpha\beta}$  are the components of the jump operator, $v_{\beta k}$ are components of the matrix of eigenvectors of the Liouvillian $M$, and $(\rho_{0})_\mu$ is the $\mu$th  component of the steady state density matrix. $N$ is the number of populations in the density matrix and $N_M$ is the dimension of $M$. In our GME calculation was $N_M=117$ and in the DME $N_M=21$ and in both cases $N=21$.

Figure \ref{Fano_U-V} shows the Fano factor $F(0)$ as a function of the charging-energy difference $\delta_U/T_C$. 
Two zones of super-Poissonian behaviour are observed: in the region near the dark state $\delta_U = 0$ and also for large charging-energy difference $\delta_U/T_C\gg 1$.  Enhancement of the Fano factor near the dark state is expected and can be understood as a dynamical channel blockade \cite{belzigDCB} with the dark state playing the role of the weakly coupled channel. The value of the Fano factor near $\delta_U=0$ is $\approx 6$, which is roughly twice that found for the single-electron dark state \cite{004}.
The precise mechanism behind the highly super-Poissonian behaviour at large $\delta_U/T_C$ is not yet clear.  However, it is a quantum coherent phenomenon, as the Fano factor drops markedly when dephasing is included \cite{dephasingFN}.


\subsection{Difference between GME and DME approach}
For asymmetric charging energies, the current, (Fig.~\ref{cur_U-Vasym}) and the Fano factor (Fig.~\ref{Fano_U-Vasym}) of the DME calculation show the same general features as the GME results, with qualitative differences becoming more evident for small values of $T_C/\Gamma$. 
In the symmetric case (Fig.~\ref{cur_U-Vsym}), however, a qualitative difference  between the GME and DME results is observed in the region of the DS.  Both calculations show a complete current suppression at $\delta_U = 0$ but, whereas the GME shows a broad antiresonance about the point $\delta_U = 0$, the DME shows a discontinuity at this point.
The discontinuous behaviour of the symmetric DME current is a reflection of the increased degeneracy of the TQD eigenstates when all $U_{ij}$ are equal.  At this point, the dark state $|\varPsi_{0}\rangle$ resides in a degenerate subspace
with two other orthogonal states:
\begin{align}
  |\varPsi_{+}\rangle =&\frac{1}{2}\big(-|1\!\uparrow\!3\!\downarrow\!\rangle\!
  -\!|2\!\uparrow\!3\!\downarrow\!\rangle\!+\!|3\!\uparrow\!1\!\downarrow\!\rangle\!+\!|3\!\uparrow\!2\!\downarrow\!\rangle\big) \\
  |\varPsi_{-}\rangle  =&\frac{1}{2\sqrt{2}}\big(-|1\!\uparrow\! 1\!\downarrow\!\rangle\!-\!|1\!\uparrow\! 2\!\downarrow\!\rangle\! -\!|2\!\uparrow\! 2\!\downarrow\!\rangle\!-|2\!\uparrow\! 1\!\downarrow\!\rangle\big)\! \nonumber \\
  &+\!\frac{1}{\sqrt{2}}|3\!\uparrow \!3\!\downarrow\!\rangle. 
  \label{degensubspace}
\end{align}
Due to this degeneracy, one has, in principle, a freedom of choice as to which combination of these states to treat as the eigenstates to be entered into the DME machinery.  Different choices result in different results.
In obtaining the plots of Figs.~\ref{cur_U-Vsym} and \ref{Fano_U-Vsym} we chose the combinations as above since, from the GME analysis, we expect a DS to form. 
However, without this {\it a priori} knowledge, any linear combination of the above vectors appears as good as any other, and if any combination other than that given explicitly above is chosen, the dark  state will not be observed since all three eigenstates will have a finite population on QD3.
We believe that this undesirable feature of the DME should be removed by a more complete second-order master equation treatment \cite{FuMa,co1,co2,co3,co4}, in which the coherences are handled properly.
For $U_{ii}\neq U_{ij}$, ($i\neq j$) or for asymmetric charging energies as in Fig.~\ref{cur_U-Vasym},
this degeneracy is lifted, the choice of eigenstates is unique and the above problems do not occur. 

The Fano factor for symmetric charging energies calculated with the DME in Fig.~\ref{Fano_U-Vsym} is also interesting because, although it reproduces the high values at $\delta_U/T_C \gg 1$, its behaviour is Poissonian, $F(0)=1$, at small values of $\delta_U/T_C$ where the GME shows super-Poissonian behaviour.  It appears then that, for symmetric charging energies the effects of the two-electron dark state in the in the DME approach are restricted solely to the singular point  $\delta_U/T_C=0$.

\section{Finite bias}

As long as the temperature is low and all energy levels of the TQD are well within the transport window, the results of the GME in the infinite bias limit are reliable \cite{{Gur},{Gur2}}.  However, the GME is not applicable away from this situation, and in this case we employ solely the DME.  Our finite-bias calculations allow us to construct stability diagrams for the system and determine the implications of the one- and two-particle dark states for tunnel spectroscopy measurements \cite{Kouw}.

\begin{figure}[tb]
  \centering
  \includegraphics[width=0.4\textwidth]{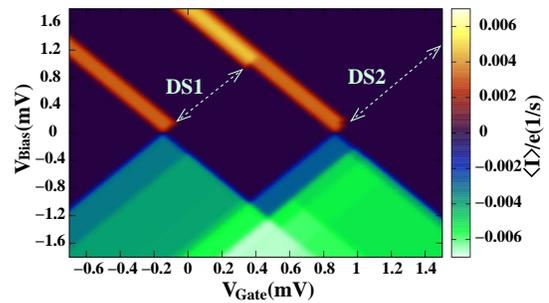}
  \caption{
  Stability diagram for the TQD with parameters such that both one- and two- electron dark states can form; in particular, $\delta_U=0$.  Here the current is plotted as a function of the source-drain bias voltage $V_{\text{Bias}}$ and gate voltage $V_{\text{Gate}}$. 
  For positive bias, most electrons enter the system on the left side where two QDs couple to the environment, and leave to the right side where one QD is coupled to the environment. Only for this bias direction can the DSs occurs. 
  The DSs are visible as breaks in the borders of the Coulomb diamonds (marked as DS1 for the one electron dark state and as DS2 for the two electron dark state); the gap near the center is due to the single-particle dark-state, that to the right, the two-particle dark state.
  The parameters used are: $U_{11}=U_{22}=U_{12}=1$meV, $U_{33}=1.2$meV, $U_{13}=U_{23}=0.95$meV, tunnel amplitude ${T_C=0.1}$meV, tunnel rate ${\Gamma=10\mu }$eV and temperature $T=150$mK.}
  \label{fig:cha_dark1}
\end{figure} 

\begin{figure}[tb]
  \centering
  \includegraphics[width=0.4\textwidth]{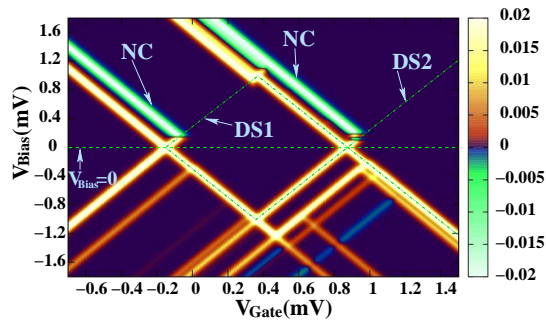}
  \caption{
  Differential conductance as a function of bias voltage $V_{\text{Bias}}$ and gate voltage $V_{\text{Gate}}$ with the same parameters as Fig.~\ref{fig:cha_dark1}.
  For $V_\text{Bias}>0$, two lines of negative differential conductance (NC) are visible, which can be attributed to the one- and two-electron DS. When a DS is hit the current decreases to zero and the differential conductance becomes negative. As in the current, the diamond pattern is broken by the formation of DSs.
  }
  \label{fig:cha_dark2}
\end{figure}

We first consider the situation with $\delta_U=0$ such that the dark state is an eigenstate of the TQD Hamiltonian.  
In accordance with the above discussion, we choose asymmetric charging energies such that methodical problems are not an issue.
The resulting current is shown in Fig.~\ref{fig:cha_dark1} and the differential conductance in Fig.~\ref{fig:cha_dark2}. 
The bias voltage $V_{\text{Bias}}$ is chosen symmetric to $V_{\text{Gate}}$ (hence $\mu_l=V_{\text{Bias}}/2 $ and $\mu_l=-V_{\text{Bias}}/2 $).
For negative bias voltage, $V_\text{Bias}<0$, the electrons enter the system through QD3 and leave it through QD1 or QD2 and the lower halves of Figs.~\ref{fig:cha_dark1} and \ref{fig:cha_dark2} show the familiar diamonds of the Coulomb blockade. 
The light-colored lines in the differential conductance diagram that lie outside  the diamonds correspond to excitation energies of the TQD.

For $V_\text{Bias}>0$, this picture changes drastically and the stability diagram is dominated by the effects of the DSs. Both one- and two-electron DSs are visible. In the current diagram they show up as breaks in the Coulomb diamonds, and in the differential conductance they give rise to clear lines of negative differential conductance (NDC). Note that, even for large $V_\text{Bias}>0$, it is possible that no current flows.
This strong asymmetry between $V_\text{Bias}<0$ and  $V_\text{Bias}>0$ demonstrates that the dark states can lead to a strong rectification of the current.
For small charging detuning $\delta_U\neq0$, the stability- and the conductance- diagrams look similar to Fig.~\ref{fig:cha_dark1} and  Fig.~\ref{fig:cha_dark2} but with a small current flowing along the line DS2.

\begin{figure}[ht!]
  \centering
  \includegraphics[width=0.45\textwidth]{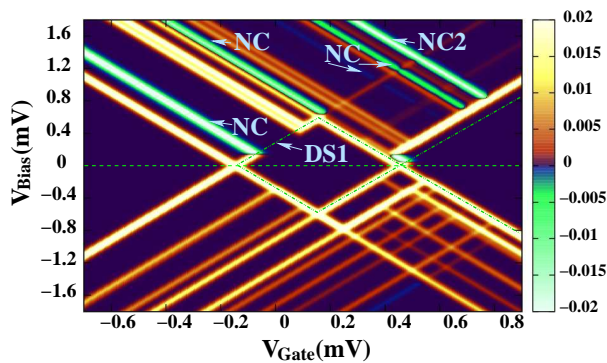}
  \caption{ 
    As Fig.~\ref{fig:cha_dark2}, but with finite charging-energy difference $\delta_U = 0.5$meV and symmetric charging energies.
    Despite the fact that the genuine two-electron dark state (\eq{Psi0}) can no longer form, lines of negative differential conductance are seen in positions similar to that in Fig.~\ref{fig:cha_dark2}.  A strong new line of NDC is also observed (marked NC2) which is related to the state of \eq{D2}.
    More structure is visible here as compared with the $\delta_U=0$ plot since degeneracies have been lifted.
 The parameters used here were $U_{ii}=1$meV, $U_{ij}=0.5$meV for $j\neq i$, ${T_C=100 \mu }$eV,  ${\Gamma=10\mu }$eV and \mbox{$T=100$mK}.}
  \label{fig:cha_U_V2}
\end{figure}

It is also of interest to consider what happens in this system far from the dark resonance.  A differential conductance diagram for this situation with symmetric charging energies is shown in Fig.~\ref{fig:cha_U_V2}. Note that in this figure, we are far from the DS such that the DME calculation should be reliable, even for symmetric charging energies.
With this set of parameters, the one-electron DS can still form and this is clearly observed as a break in the Coulomb diamond and a diagonal line of negative differential conductance.
Moreover, traces of a two-electron dark states are still visible, manifest as lines of negative differential conductance in the two-electron sector.  These occur because, although the full dark-state is no longer an eigenstate of the system Hamiltonian, there still exists related states which, although not completely dark, can only support a comparatively small current.  
In particular, there is one NDC line (marked NC2) in the two-electron region of the stability diagram for which the suppression is particularly strong.  The corresponding stationary density matrix has very low occupation on the third QD and predominantly given by $\rho_\mathrm{stat}\approx\op{\varPsi_{D2}}{\varPsi_{D2}}$ with
\begin{eqnarray}
 |\varPsi_{D2}\rangle=\frac{1}{\sqrt{2}}\rb{d^\dag_{1\uparrow}d^\dag_{1\downarrow}-d^\dag_{2\uparrow}d^\dag_{2\downarrow}}\ket{0}. \label{eq_dark_state2}
  \label{D2}
\end{eqnarray}
This state is similar to the dark state of \eq{Psi0} but with only contributions with two electrons in each dot.
With an increasing difference between $U_{ii}$ and $U_{ij}$ for $j\neq i$ the line of negative differential conductance which corresponds to NC2 becomes stronger and also the weight of the state \eq{eq_dark_state2} increases.

\section{Conclusions}

We have shown in this article that the TQD in transport can exhibit current suppression due to a two-particle DS.  This represents an extension of the single-electron DS concept familiar from optics to the multiple-electron domain. In the TQD, the two-electron DS can exist only under the condition that the charging energies $U_{11}=U_{22}=U_{12}$ are equal.  
Importantly though, even if these energies are not equal, traces of the DS still remain, and should be observable in experiment. 
 
Specifically, these traces are a large Fano factor near the DS and pronounced lines of negative differential  
\mbox{conductance} in the stability diagrams.
We note that an optical two-electron DS was presented in Ref.~\cite{opDDS}. An analogous mesoscopic transport configuration would not lead to a DS formation and hence current blockade, since the dark-state has finite occupation of all states in the localized basis.

On a technical level, we have compared the GME and DME approach in the infinite bias limit.  We believe that the gap between these two methods can be effectively bridged with a more complete second-order master equation treatment, e.g.  \cite{FuMa,co1,co2,co3,co4}. This should enable us to calculate transport properties at finite bias without having to concern ourselves with the discontinuity exhibited by the DME.

Further work includes investigation of dark states with still higher electron numbers and the influence of magnetic field.

\section*{Acknowledgment}
We are grateful to R.~Aguado, B. R.~Bu\l ka, M.~Busl, G.~Platero and F.~Renzoni for useful discussions.

\appendix
\section{Opposite-spin two-electron basis}
\label{appen_1}
In the basis
$\left\{
\ket{1\!\!\uparrow\! 1 \!\!\downarrow\!}, 
\ket{1\!\!\uparrow\! 2 \!\!\downarrow\!}, 
\ket{1\!\!\uparrow\! 3 \!\!\downarrow\!},
\ket{2\!\!\uparrow\! 2 \!\!\downarrow\!}, 
\ket{2\!\!\uparrow\! 1 \!\!\downarrow\!},
\ket{2\!\!\uparrow\! 3 \!\!\downarrow\!},
\right.$
$\left.
\ket{3\!\!\uparrow\! 3 \!\!\downarrow\!}, 
\ket{3\!\!\uparrow\! 1 \!\!\downarrow\!},
\ket{3\!\!\uparrow\! 2 \!\!\downarrow\!}\right\}
$, the Hamiltonian for two electrons of opposite spins reads
\begin{widetext}
\begin{align}
H_{\uparrow\downarrow} =& \begin{pmatrix}
2E_1\!\!+\!\!U_{11} &0                             &T_C                             &0                  &0 
&0                             &0                  &T_C                             &0 \\
0                  &E_1\!\!+\!\!E_2\!\!+\!\!U_{12} &T_C                             &0                  &0 
&0                             &0                  &0                             &T_C \\
T_C                  &T_C                             &E_1\!\!+\!\!E_3\!\!+\!\!U_{13} &0                  &0 
&0                             &T_C                  &0                             &0 \\
0                  &0                             &0                             &2E_2\!\!+\!\!U_{22} &0 
&T_C                             &0                  &0                             &T_C \\
0                  &0                             &0                             &0                  &E_2\!\!+\!\!E_1\!\!+\!\!U_{12} 
&T_C                             &0                  &T_C                             &0 \\
0                  &0                             &0                             &T_C                  &T_C 
&E_2\!\!+\!\!E_3\!\!+\!\!U_{23} &T_C                  &0                             &0 \\
0                  &0                             &T_C                             &0                  &0 
&T_C                             &2E_3\!\!+\!\!U_{33} &T_C                             &T_C \\
T_C                  &0                             &0                             &0                  &T_C 
&0                             &T_C                  &E_3\!\!+\!\!E_1\!\!+\!\!U_{13} &0 \\
0                  &T_C                             &0                             &T_C                  &0 
&0                             &T_C                  &0                             &E_3\!\!+\!\!E_2\!\!+\!\!U_{23} \end{pmatrix} 
.
\end{align}


\section{Tunnel operators} \label{appen_2}
The second-quantized operators appearing in Eq.\eqref{auh} can be written in the relevant many-body basis as:
\begin{eqnarray}
d_{3 \uparrow} &=&\op{0}{3\!\! \uparrow\!\!}-\op{1\!\! \uparrow}{1\!\! \uparrow\!3\!\! \uparrow\!\!}-\op{2\!\! \uparrow}{2\!\! \uparrow\!3\!\! \uparrow\!\!}+\op{1\!\! \downarrow}{3\!\! \uparrow\!1\!\! \downarrow\!\!}+\op{2\!\! \downarrow}{3\!\! \uparrow\!2\!\! \downarrow\! \!}+\op{3\!\! \downarrow}{3\!\! \uparrow\!3\!\! \downarrow\!\!},
\\
d_{3 \downarrow} &=&\op{0}{3\!\! \downarrow\!\!}-\op{1\!\! \downarrow}{1\!\! \uparrow\!3\!\! \downarrow\!\!}-\op{2\!\! \uparrow}{2\!\! \uparrow\!3\!\! \downarrow\!\!}-\op{3\!\! \uparrow}{3\!\! \uparrow\!3\!\! \downarrow\!\!}-\op{1\!\! \downarrow}{1\!\! \downarrow\!3\!\! \downarrow\!\!}-\op{2\!\! \downarrow}{2\!\! \downarrow\!3\!\! \downarrow\! \!},
\\
d_{1 \uparrow}^\dagger &=&\op{1\!\! \uparrow}{0}
 +\op{1\!\! \uparrow\!2\!\! \uparrow}{2\!\! \uparrow\!\!}+\op{1\!\! \uparrow\!3\!\! \uparrow}{3\!\! \uparrow\!\!}
 +\op{1\!\! \uparrow\!1\!\! \downarrow}{1\!\! \downarrow\!\!}+\op{1\!\! \uparrow\! 2\!\! \downarrow}{2\!\! \downarrow\!\!}+\op{1\!\! \uparrow\! 3\!\! \downarrow}{3\!\! \downarrow\!\!}
\\
d_{2 \uparrow}^\dagger &=&\op{2\!\! \uparrow}{0}
 -\op{1\!\! \uparrow\!2\!\! \uparrow}{1\!\! \uparrow\!\!}+\op{2\!\! \uparrow\!3\!\! \uparrow}{3\!\! \uparrow\!\!}
 +\op{2\!\! \uparrow\!1\!\! \downarrow}{1\!\! \downarrow\!\!}+\op{2\!\! \uparrow\! 2\!\! \downarrow}{2\!\! \downarrow\!\!}
 +\op{2\!\! \uparrow\!3\!\! \downarrow}{3\!\! \downarrow\!\!},
\\
d_{1 \downarrow}^\dagger &=&\op{1\!\! \downarrow}{0}
-\op{1\!\! \uparrow\!1\!\! \downarrow}{1\!\! \uparrow\!\!}-\op{2\!\! \uparrow\! 1\!\! \downarrow}{2\!\! \uparrow\!\!}
-\op{3\!\! \uparrow\!1\!\! \downarrow}{3\!\! \uparrow\!\!} 
 +\op{1\!\! \downarrow\!2\!\! \downarrow}{2\!\! \downarrow\!\!}+\op{1\!\! \downarrow\!3\!\! \downarrow}{3\!\! \downarrow\!\!},
\\
d_{2 \downarrow}^\dagger &=&\op{2\!\! \downarrow}{0}
-\op{1\!\! \uparrow\!2\!\! \downarrow}{1\!\! \uparrow\!\!}-\op{2\!\! \uparrow\! 2\!\! \downarrow}{2\!\! \uparrow\!\!}
-\op{3\!\! \uparrow\!2\!\! \downarrow}{3\!\! \uparrow\!\!} 
 -\op{1\!\! \downarrow\!2\!\! \downarrow}{1\!\! \downarrow\!\!}+\op{2\!\! \downarrow\!3\!\! \downarrow}{3\!\! \downarrow\!\!}.
\end{eqnarray}

\end{widetext}



\begin{thebibliography}{99}
\bibitem{dephasingFN} The dephasing due to background charge fluctuations was implemented along the lines of Refs.~\cite{003,005,Gur3}.
\end{thebibliography}
\end{document}